\documentclass[aoas,nameyear,noautosecdot,dvips]{arximspdf}
\usepackage{graphics}

\doi{10.1214/09-AOAS267}
\volume{3}
\issue{4}
\pubyear{2009}
\firstpage{1805}
\lastpage{1830}

\makeatletter
\newtheorem{definition}{Definition}

\newcommand{\bd}{\begin{definition}}
\newcommand{\ed}{\end{definition}}
\newcommand{\bi}{\begin{itemize}}
\newcommand{\ei}{\end{itemize}}

\makeatother

\begin{document}
\begin{frontmatter}

\title{Smoothed ANOVA with spatial effects as a~competitor to MCAR in
multivariate spatial~smoothing}
\runtitle{SANOVA and MCAR}

\begin{aug}
\author[A]{\fnms{Yufen} \snm{Zhang}},
\author[B]{\fnms{James S.} \snm{Hodges}} \and
\author[C]{\fnms{Sudipto} \snm{Banerjee}\thanksref{t1}\ead[label=e1]{sudiptop@biostat.umn.edu}\corref{}}
\runauthor{Y. Zhang, J. S. Hodges and S. Banerjee}
\affiliation{Novartis Pharmaceuticals,  University of Minnesota and University of Minnesota}
\thankstext{t1}{Supported in part by NIH Grant 1-R01-CA95995.}
\address[A]{Y. Zhang\\
Novartis Pharmaceuticals\\
East Hanover, New Jersey 07936\\
USA} 
\address[B]{J. S. Hodges\\
Division of Biostatistics\\
School of Public Health\\
University of Minnesota\\
Minneapolis, Minnesota 55455\\
USA}
\address[C]{S. Banerjee\\
Division of Biostatistics\\
School of Public Health\\
University of Minnesota\\
Minneapolis, Minnesota 55455\\
USA\\
\printead{e1}}
\end{aug}
%
\received{\smonth{8} \syear{2008}}
\revised{\smonth{12} \syear{2008}}

\begin{abstract}
Rapid developments in geographical information systems (GIS)
continue to generate interest in analyzing complex spatial datasets. One area
of activity is in creating smoothed disease maps to describe the geographic
variation of disease and generate hypotheses for apparent differences in risk.
With multiple diseases, a multivariate conditionally autoregressive (MCAR)
model is often used to smooth across space while accounting for associations
between the diseases. The MCAR, however, imposes complex covariance structures
that are difficult to interpret and estimate. This article develops a much
simpler alternative approach building upon the techniques of smoothed ANOVA
(SANOVA).
Instead of simply shrinking effects without any structure, here we use SANOVA
to smooth spatial random effects by taking advantage of the spatial structure.
We extend SANOVA to cases in which one factor is a spatial lattice, which is
smoothed using a CAR model, and a second factor is, for example, type of
cancer. Datasets routinely lack enough information to identify the additional
structure of MCAR. SANOVA offers a simpler and more intelligible structure
than the MCAR while performing as well. We demonstrate our approach with
simulation studies designed to compare SANOVA with different design matrices
versus MCAR with different priors. Subsequently a cancer-surveillance dataset,
describing incidence of 3-cancers in Minnesota's 87 counties, is analyzed
using both approaches, showing the competitiveness of the SANOVA approach.
\end{abstract}


\end{frontmatter}
%
\section{\texorpdfstring{Introduction.}{Introduction}}\label{C:intro}
Statistical modeling and analysis of spatially referenced data receive
considerable interest due to the increasing availability of geographical
information systems (GIS) and spatial databases. For data aggregated over
geographic regions such as counties, census tracts or ZIP codes (often called
\textit{areal} data), with individual identifiers and precise
locations removed,
inferential objectives focus on models for spatial clustering and variation.
Such models are often used in epidemiology and public health to understand
geographical patterns in disease incidence and morbidity. Recent
reviews of
methods for such data include Lawson et al. (\citeyear{Lawson99}),
Elliott et
al. (\citeyear{Elliott00}), Waller
and Gotway (\citeyear{Waller04}) and Rue and Held (\citeyear{Rue05}).
Traditionally such data have been modeled using conditionally specified
probability models that shrink or smooth spatial effects by borrowing strength
from neighboring regions. Perhaps the most pervasive model is the conditionally
autoregressive (CAR) family pioneered by Besag (\citeyear{Besag74}),
which has been
widely investigated and applied to spatial epidemiological data
[Wakefield~(\citeyear{Wakefield07}) gives
an excellent review]. Recently the CAR has been extended to multivariate
responses, building on multivariate conditional autoregressive (MCAR) models
described by Mardia (\citeyear{Mardia88}). Gelfand and Vonatsou
(\citeyear{Gelfand03})
and Carlin and Banerjee~(\citeyear{Carlin03}) discussed their use in
hierarchical models, while Kim, Sun and Tsutakawa
(\citeyear{Kim01}) presented a different ``twofold CAR'' model for
counts of two
\mbox{diseases} in
each areal unit. Other extensions allowing flexible modeling of
cross-correlations include Sain and Cressie (\citeyear{Sain02}), Jin,
Carlin and
Banerjee (\citeyear{Jin05}) and
Jin, Banerjee and Carlin (\citeyear{Jin07}).
The MCAR can be viewed as a conditionally specified probability model for
interactions between space and an attribute of interest. For instance, in
disease mapping interest often lies in modeling geographical patterns in
disease rates or counts of several diseases. The MCAR acknowledges dependence
between the diseases as well as dependence across space. However, practical
difficulties arise from MCAR's elaborate dependence structure: most interaction
effects will be weakly identified by the data, so the dependence
structure is
poorly identified. In hierarchical models [e.g., Gelfand and Vonatsou
(\citeyear{Gelfand03}), Jin, Carlin and Banerjee (\citeyear
{Jin05,Jin07})], strong prior distributions may improve
identifiability, but
this is not uncontroversial, as inferences are sensitive to the prior
and perhaps
unreliable without genuine prior information.
This article proposes a~much simpler and more interpretable alternative
to the MCAR, modeling multivariate spatial effects using smoothed
analysis of variance (SANOVA) as developed by Hodges, Carlin and Fan (\citeyear
{Hodges07}),
henceforth HCSC. Unlike an ANOVA that is used to
identify some interaction effects to retain and others to remove,
SANOVA mostly retains effects that are large, mostly removes those that
are small, and partially retains middling effects. (Loosely speaking,
``large,'' ``middling'' and ``small'' describe the size of the unsmoothed
effects compared to their standard errors.) To accommodate rich
dependence structures, MCAR introduces weakly identifiable parameters
that complicate estimation. SANOVA, on the other hand, focuses instead
on smoothing interactions to yield more stable and reliable results.
Our intended contribution is to show how SANOVA can solve the multiple
disease mapping problem while avoiding the dauntingly complex
covariance structures imposed by MCAR and its generalizations. We
demonstrate that SANOVA produces inference that is largely
indistinguishable from MCAR, yet SANOVA is simpler, more explicit,
easier to put priors on and easier to estimate.
The rest of the article is as follows. Section \ref{sec2} reviews
SANOVA and MCAR,
identifying SANOVA as a special case of MCAR. Section \ref{S:simulation1} is a
``tournament'' of
simulation experiments comparing SANOVA with MCAR for normal and
Poisson data,
while Section \ref{sec4} analyzes data describing the number of deaths from
lung, larynx and esophagus cancer in Minnesota between 1990 and 2000. A
summary and discussion of future research in Section~\ref{discussion}
concludes the
paper. Zhang, Hodges and Banerjee (\citeyear{Zhang09}) (Appendices)
gives computational and
technical details.

\section{\texorpdfstring{The competitors.}{The competitors}}\label{sec2}
%
\subsection{\texorpdfstring{Smoothing spatial effects using SANOVA.}{Smoothing spatial effects using SANOVA}}
%
\subsubsection{\texorpdfstring{SANOVA for balanced, single-error-term models [HCSC (\protect\citeyear{Hodges07})].}
{SANOVA for balanced, single-error-term models [HCSC (2007)]}}\label{S:balance}
Consider a balanced, single-error-term analysis of variance, with $M_1$ degrees
of freedom for main effects and $M_2$ degrees of freedom for interactions.
Specify this ANOVA as a linear model: let $A_1$ denote columns in the design
matrix for main effects, and $A_2$ denote columns in the design matrix for
interactions. Assume the design has $c$ cells and $n$ observations per cell,
giving $cn$ observations in total. To simplify later calculations,
normalize the
columns of $A_1$ and $A_2$ so ${A_1'}A_1=I_{M_1}$ and ${A_2'}A_2=I_{M_2}$.
(Note: HCSC normalized columns differently, fixing
${A_1'}A_1=cnI_{M_1}$ and
${A_2'}A_2=cnI_{M_2}$.) Then write the ANOVA as
%
\begin{equation}{\label{datacase}}
\mathbf{y} = [A_1|A_2]
\left[\matrix{
\bolds{\Theta}_1\cr
\bolds{\Theta}_2}\right]
+\bolds{\epsilon}= A_1\bolds{\Theta}_1 +
A_2\bolds{\Theta}_2 + \bolds{\epsilon},
\end{equation}
where $\bolds{\epsilon}\sim N(0,\frac{1}{\eta_0}I)$ with $\eta_0$
being a
precision, $\mathbf{y}$ is $cn \times1$, $A_1$ is $cn \times M_1$,
$A_2$ is
$cn \times M_2$, $\bolds{\Theta}_1$ is $M_1 \times1$, $\bolds
{\Theta
}_2$ is $M_2 \times1$, and $\bolds{\epsilon}$ is $cn \times1$.
This ANOVA is smoothed by further modeling $\bolds{\Theta}$. HCSC emphasized
smoothing interactions, although main effects can be smoothed by
exactly the
same means. Following HCSC, we add constraints (or a prior) on
$\bolds{\Theta}_2$ as
$\theta_{ M_1+ j}\sim N(0,1/\eta_j)$ for
$j=1,\ldots,M_2 $, written as
%
\begin{equation}{\label{constraintcase}}
\mathbf{0}_{M_2} = \bigl[0_{M_2 \times M_1}|I_{M_2}\bigr]
\left[\matrix{
\bolds{\Theta}_1\cr
\bolds{\Theta}_2}
\right] +\bolds{\delta},
\end{equation}
where $\bolds{\delta}\sim N(0, \operatorname{diag}(\frac{1}{\eta_j}))$, in the
manner of
Lee and
Nelder (\citeyear{Lee96}) and Hodges (\citeyear{Hodges98}). Combining
(\ref{datacase}) and
(\ref{constraintcase}), express this hierarchical model as a linear model:
%
\begin{equation}{\label{dataconstraint}}
\left[\matrix{
\mathbf{y}\cr\mathbf{0}_{M_2}}
\right]=\left[\matrix{
A_1 &
A_2\cr
0_{M_2 \times M_1}& I_{M_2}}
\right]
\left[\matrix{
\bolds{\Theta}_1 \cr \bolds{\Theta}_2}
\right]+
\left[
\matrix{\bolds{\epsilon}\cr \bolds{\delta}}
\right].
\end{equation}
More compactly, write
%
\begin{equation}{\label{olm}}
\mathbf{Y}=X\bolds{\Theta}+ \mathbf{e},
\end{equation}
where $\mathbf{Y}$ has dimension $(cn+M_2) \times1$ and
$\mathbf{e}$'s
covariance $\Gamma$ is block diagonal with blocks
$\Gamma_1=\frac{1}{\eta_0}I_{cn}$ for the data cases (rows of $X$
corresponding
to the observation~$\mathbf{y}$) and $\Gamma_2 =
\operatorname{diag}({1}/{\eta_1},\ldots,{1}/{\eta_{ M_2}})$ for the
constraint cases (rows of $X$ with error term $\bolds{\delta}$). For
convenience, define the matrix $X_D=[A_1|A_2]$, the data-case part of $X$.
This development can be done using the mixed linear model (MLM)
formulation traditionally written as $\mathbf{y} = X\bolds{\beta} +
Z\mathbf{u} + \bolds{\epsilon}$, where our (\ref{datacase}) supplies
this equation and $\mathbf{u} = \bolds{\Theta}_2 \sim N(\mathbf{0},
\Gamma_2)$. The development to follow can also be done using the MLM
formulation at the price of slightly greater complexity, so we omit it.
HCSC developed SANOVA for exchangeable priors on groups formed from components
of $\bolds{\Theta}_2$. The next section develops the extension to
spatial smoothing.

\subsubsection{\texorpdfstring{What is CAR?}{What is CAR?}}\label{Q}
Suppose a map has $N$ regions, each with an unknown quantity of interest
$\phi_i$, $i=1,\ldots,N$. A conditionally autoregressive (CAR) model specifies
the full conditional distribution of each $\phi_i$ as
%
\begin{equation}\label{psicond}
\phi_i \mid\phi_j, j\neq i,
\sim N \Biggl(\frac{\alpha}{m_i}\sum_{i \sim j}\phi_j,
\frac{1}{\tau m_i}\Biggr),\qquad i, j =1, \ldots, N ,
\end{equation}
where $i \sim j$ denotes that region $j$ is a {\em neighbor} of region $i$
(typically defined as spatially adjacent), and $m_i$ is the number of region
$i$'s neighbors. Equation (\ref{psicond}) reduces to the well-known intrinsic
conditionally autoregressive (ICAR) model [Besag, York and Molli\'{e} (\citeyear
{Besag91})] if
$\alpha=1$ or
an independence model if $\alpha=0$. The ICAR model induces ``local'' smoothing
by borrowing strength from neighbors, while the independence model assumes
spatial independence and induces ``global'' smoothing. The CAR prior's smoothing
parameter $\alpha$ also controls the strength of spatial dependence among
regions, though it has long been appreciated that a fairly large
$\alpha
$ may be
required to induce large spatial correlation; see Wall (\citeyear
{Wall04}) for recent
discussion and examples.
It is well known [e.g., Besag (\citeyear{Besag74})] that the conditional
specifications in
(\ref{psicond}) lead to a valid joint distribution for $\bolds{\phi
} =
(\phi_1,\ldots,\phi_N){'}$ expressed in terms of the map's neighborhood
structure. If $Q$ is an $N \times N$ matrix such that $Q_{ii}=m_i$,
$Q_{ij} = -
\alpha$ whenever $i \sim j$ and $Q_{ij} = 0$ otherwise, then the
intrinsic CAR
model [Besag, York and Molli\'{e} (\citeyear{Besag91})] has density
%
\begin{eqnarray}{\label{car}}
p(\bolds{\phi}|\tau) &\sim&\tau^{{N^*}/{2}}\operatorname{exp} \biggl(-\frac{\tau}{2}
\bolds{\phi}'Q\bolds{\phi}\biggr),\qquad \mbox{with}
\nonumber
\\[-8pt]
\\[-8pt]
\nonumber
N^* &=& \cases{
N, &\quad $\mbox{if } \alpha\in(0,1)$, \cr
N-G, &\quad $\mbox{if } \alpha=1$.}
\end{eqnarray}
In (\ref{car}) $\tau$ is the spatial precision parameter, $\tau Q$ is
the precision matrix in this multivariate normal distribution and $G$
is the number of ``islands'' (disconnected parts) in the spatial map
[Hodges, Carlin and Fan (\citeyear{Hodges03})]. When $\alpha\in(0,1)$, (\ref
{car}) is a~proper
multivariate normal distribution. When $\alpha=1$, $Q$ is singular with
$Q\mathbf{1}= \mathbf{0}$; $Q$ has rank $N-G$ in a map with $G$ islands,
therefore, the exponent on $\tau$ becomes $(N-G)/2$.
In hierarchical models, the CAR model is usually used as a prior on spatial
random effects. For instance, let $Y_i$ be the observed number of cases
of a~disease in region $i$, $i=1, \ldots, N$, and $E_i$ be the expected
number of
cases in region $i$. Here the $Y_i$ are treated as random variables,
while the
$E_i$ are treated as fixed and known, often simply proportional to the number
of persons at risk in region $i$. For rare diseases, a Poisson
approximation to
a binomial sampling distribution for disease counts is often used, so a commonly
used likelihood for mapping a single disease is
%
\begin{equation}\label{pois1}
Y_{i} \stackrel{\mathrm{ind}} \sim
\operatorname{Poisson}(E_{i}e^{\mu_{i}}),\qquad i=1, \ldots, N,
\end{equation}
where $\mu_{i}=\mathbf{x}_i'\bolds{\beta}+ \phi_i$. The $\mathbf
{x}_i$ are explanatory,
region-specific regressors with coefficients $\bolds{\beta}$ and
the parameter $\mu_i$ is the log-relative risk describing departures
of observed
from expected counts, that is, from $E_i$. The hierarchy's next level is
specified by assigning the CAR distribution to $\bolds{\phi}$ and a
hyper-prior to the spatial precision parameter $\tau$. In the hierarchical
setup, the improper ICAR with $\alpha=1$ gives proper posterior distributions
for spatial effects. In practice, Markov chain Monte Carlo (MCMC)
algorithms are
designed for estimating posteriors from such models and the appropriate number
of linear constraints on the $\phi$ suffices to ensure sampling from
proper posterior distributions [Banerjee, Carlin and Gelfand~(\citeyear
{Banerjee04}),
pages 163--164, give details].

\subsubsection{How does CAR fit into SANOVA?}
To use CAR in SANOVA, the key is re-expressing the improper CAR, that
is, (\ref{car}) with $\alpha=1$. Let
$Q$ have spectral decomposition $Q=VDV'$, where $V$ is an orthogonal
matrix with
columns containing $Q$'s eigenvectors and $D$ is diagonal with nonnegative
diagonal entries. $D$ has~$G$ zero diagonal entries, one of which
corresponds to
the eigenvector $\frac{1}{\sqrt{N}}\mathbf{1}_N$, by convention the $N${th}
(right-most) column in $V$. Define a new parameter
$\bolds{\Theta}=V'\bolds{\phi}$, so $\bolds{\Theta}$ has
dimension $N$ and precision matrix $\tau D$. Giving an $N$-vector
$\bolds{\Theta}$ a normal prior with mean zero and precision $\tau
D$ is
equivalent to giving\vspace*{1.5pt} $\bolds{\phi}=V\bolds{\Theta}$ a~CAR
prior with
precision $\tau Q$. $\bolds{\Theta}$ consists of
$\Theta_N=\frac{1}{\sqrt{N}}\mathbf{1}_N'\bolds{\phi}=\sqrt
{N}\mbox{ }
\overline{\bolds{\phi}}$, the scaled average of the $\bolds{\phi}_i$,
along with $N-1$ contrasts in $\bolds{\phi}$, which are orthogonal to
$\frac{1}{\sqrt{N}}\mathbf{1}_N$ by construction. Thus, the CAR prior
is informative
(has positive precision) only for contrasts in $\bolds{\phi}$, while
putting zero precision on $\Theta_{\mathit{GM}} = \Theta_N =
\frac{1}{\sqrt{N}}\mathbf{1}'_N \bolds{\phi}$, the overall
level, and on
$G-1$ orthogonal contrasts in the levels of the $G$ islands. In other
words, the CAR model can be thought of as a prior distribution on the
contrasts rather than individual effects (hence the need for the
sum-to-zero constraint). A~related result, discussed in Besag, York and Molli\'{e}
(\citeyear{Besag95}), shows the CAR to be a member of a family of ``pairwise
difference'' priors.
This reparameterization allows the CAR model to fit into the ANOVA framework,
with $\Theta_{\mathit{GM}}$ corresponding to the ANOVA's grand mean and the
rest of
$\bolds{\Theta}$, $\bolds{\Theta}_{\mathit{Reg}}$, corresponding to ${V^{(-
)}}{}'\bolds{\phi}$, where $V^{(-)}$ is $V$ excluding the column
$\frac{1}{\sqrt{N}}\mathbf{1}_N$, consisting of $N-1$ orthogonal contrasts
among the $N$ regions and giving the $N-1$ degrees of freedom in the usual
ANOVA:
\begin{eqnarray*}
\bolds{\phi}&=& [
{\phi}_1, {\phi}_2, \ldots, {\phi}_N]'\\
& = & V \bolds{\Theta}\\
&=& \left[\matrix{ V^{(-)}&\tfrac{1}{\sqrt{N}}\mathbf{1}_{N}}\right] \left[\matrix{
 \bolds{\Theta}_{\mathit{Reg}}\cr
\Theta_{\mathit{GM}}}\right].
\end{eqnarray*}
Giving $\bolds{\phi}$ a CAR prior is equivalent to giving
$\bolds{\Theta}$ a $N(\mathbf{0}, \tau D)$ prior; the latter are the
``constraint
cases'' in HCSC's SANOVA structure. The precision $D_{NN}=0$ for the
overall level
is equivalent to a flat prior on $\Theta_{\mathit{GM}}$, though $\Theta_{\mathit{GM}}$ could
alternatively have a normal prior with mean zero and finite variance.
If $G>1$,
the CAR prior also puts zero precision on $G-1$ contrasts in
$\bolds{\phi}$, which are contrasts in the levels of the $G$ islands
[Hodges, Carlin and Fan (\citeyear{Hodges03})].

\subsection{\texorpdfstring{SANOVA as a competitor to MCAR.}{SANOVA as a competitor to MCAR}}\label{S:sanovavsmcar}
\subsubsection{\texorpdfstring{Multivariate conditionally autoregressive (MCAR)
models.}{Multivariate conditionally autoregressive (MCAR)
models}}\label{ss:mcar}
With multiple diseases, we have unknown $\phi_{ij}$ corresponding to
region $i$
and disease $j$, where $i=1,\ldots,N$ and $j=1,\ldots,n$. Letting
$\Omega$ be a
common precision matrix (i.e., inverse of the covariance matrix) representing
correlations between the diseases in a given region, MCAR distributions arise
through conditional specifications for $\bolds{\phi}_i =
(\phi_{i1},\ldots,\phi_{in})'$:
%
\begin{equation}\label{mcarcond}
\bolds{\phi}_i|\{\bolds{\phi}_{i'}\}_{i'\neq i} \sim
\operatorname{MVN}\biggl(\frac{\alpha}{m_i}\sum_{i'\sim
i}\bolds{\phi}_{i'},\frac{1}{m_i}\Omega^{-1}\biggr).
\end{equation}
These conditional distributions yield a joint distribution for
$\bolds{\phi}=(\bolds{\phi}{}'_1,\ldots,\bolds{\phi}{}'_N)'$:
%
\begin{equation}{\label{mcar}}
f(\bolds{\phi}|\Omega) \propto\|\Omega\|^{(N-G)/{2}}
\operatorname{exp}
{\bigl(-
\tfrac{1}{2}\bolds{\phi}'(Q \otimes
\Omega)\bolds{\phi}\bigr)},
\end{equation}
%
where $Q$ is defined as in Section \ref{Q} and again
(\ref{mcar}) is an improper density when $\alpha=1$. However, as for the
univariate CAR, this yields proper posteriors in conjunction with a proper
likelihood.
The specification above is a ``separable'' dispersion structure, that
is, the
covariances between the diseases are invariant across regions. This may seem
restrictive, but relaxing this restriction gives even more complex dispersion
structures [see Jin, Banerjee and Carlin (\citeyear{Jin07}) and
references therein]. As
mentioned earlier, our focus is to retain the model's simplicity without
compromising the primary inferential goals. We propose to do this using SANOVA
and will compare it with the separable MCAR only.

\subsubsection{\texorpdfstring{SANOVA with Minnesota counties as one
factor.}{SANOVA with Minnesota counties as one
factor}}\label{ss:sanovamcar}
We now describe the SANOVA model using the Minnesota 3-cancer dataset. Consider
the Minnesota map with $N=87$ counties, and suppose each county has
counts for
$n=3$ cancers. County $i$ has an $n$-vector of parameters describing
the $n$
cancers, $\bolds{\phi}_i = (\phi_{i1},\phi_{i2},\ldots,\phi_{in})'$;
define the $Nn$ vector $\bolds{\phi}$ as $\bolds{\phi} =({\bolds{\phi}_1'}, {\bolds{\phi}_2'},\ldots,
{\bolds{\phi}_N'})'$. For now, we are vague about the specific
interpretation of ${\phi}_{ij}$; the following description applies to
any kind
of data. Assume the $N \times N$ matrix $Q$ describes neighbor pairs among
counties as before.
The SANOVA model for this problem is a 2-way ANOVA with factors cancer
(``$\mathit{CA}$,'' $n$ levels) and county (``$\mathit{CO}$,'' $N$ levels) and no
replication. As
in Section \ref{S:balance}, we model $\bolds{\phi}$ with a
saturated linear model and put the grand mean and the main effects in
their traditional positions as in ANOVA (matrix dimensions and
definitions appear below the equation):
%
\begin{eqnarray}\label{long}
\bolds{\phi}&=& [
\bolds{\phi}_1', \bolds{\phi}_2', \ldots, \bolds{\phi}_N']'= [ A_1|A_2 ] \bolds{\Theta}\nonumber
\\
&=&\left[\vphantom{\frac{1}{\sqrt{Nn}}\mathbf{1}_{Nn}}\right.\hspace*{115pt}\left|\vphantom{\frac{1}{\sqrt{Nn}}\mathbf{1}_{Nn}}\right.
\\[-28pt]
&&\hspace*{4pt}\matrix{
\mathop{\underbrace{\tfrac{1}{\sqrt{Nn}}\mathbf{1}_{Nn}}_{\mathrm{Grand\mbox{ }mean}}}\limits_{Nn\times1}&
\mathop{\mathop{\underbrace{\tfrac{1}{\sqrt{N}}\mathbf
{1}_{N} \otimes H_{\mathit{CA}}}_{\mathrm{Cancer}}}\limits_{\mathrm{main}\
\mathrm{effect}}}\limits_{Nn\times(n-1)}}\nonumber
\\ 
&&\hspace*{5.86pt}\matrix{\mathop{\mathop{\underbrace{ V^{(-)} \otimes
\tfrac{1}{\sqrt{n}}\mathbf{1}_n}_{\mathrm{County}}}\limits_{\mathrm{main}\ \mathrm{effect}}}\limits_{Nn\times(N-1)} & \mathop{\mathop{\underbrace{
V^{(-)} \otimes
{H^{(1)}_{\mathit{CA}}}\ \ \cdots\ \ V^{(-)}
\otimes{H^{(n-1)}_{\mathit{CA}}}}_{\mathrm{Cancer}\times\mathrm{County}}}\limits_{\mathrm{interaction}}}\limits_{Nn\times(N-1)(n-1)}}
\nonumber\\[-61pt]
&&\hspace*{225pt}\left.\vphantom{\frac{1}{\sqrt{Nn}}\mathbf{1}_{Nn}}\right]\nonumber
\\[38pt]
&&{}\times \left[\matrix{
\Theta_{\mathit{GM}}\vspace*{2pt}\cr
\bolds{\Theta}_{\mathit{CA}}\vspace*{2pt} \cr
\bolds{\Theta}_{\mathit{CO}} \vspace*{2pt}\cr
\bolds{\Theta}_{{\mathit{CO}\times \mathit{CA}}}}\right],\nonumber
\end{eqnarray}
where $H_{\mathit{CA}}$ is an $n \times(n-1)$ matrix whose columns are
contrasts among cancers, so $\mathbf{1}_n'H_{\mathit{CA}}=\mathbf{0}'_{n-1}$, and
$H_{\mathit{CA}}'H_{\mathit{CA}}=I_{n-1}$; ${H^{(j)}_{\mathit{CA}}}$ is the $j${th} column of $H_{\mathit{CA}}$;
and $V^{(-)}$ is $V$ without its $N${th} column $\frac{1}{\sqrt
{N}}\mathbf{1}_N$, so it has $N-1$ columns, each a contrast among
counties, that is,
$\mathbf{1}_N' V^{(-)}=\mathbf{0}'_{N-1}$, and $V^{(-)\prime}V^{(-)}=I_{N-1}$.
The column labeled ``Grand mean'' corresponds to the ANOVA's grand mean
and has
parameter $\Theta_{\mathit{GM}}$; the other blocks of columns labeled as main
effects and
interactions correspond to the analogous ANOVA effects and to their respective
parameters
$\bolds{\Theta}_{\mathit{CA}},\bolds{\Theta}_{\mathit{CO}},\bolds{\Theta}_{\mathit{CO} \times
\mathit{CA}}$.
Defining prior distributions on $\bolds{\Theta}$ completes the SANOVA
specification. We put independent flat priors (normal with large
variance) on
$\Theta_{\mathit{GM}}$ and $\bolds{\Theta}_{\mathit{CA}}$, which are, therefore, not
smoothed. This is equivalent to putting a flat prior on each of the $n$
cancer-specific means. To specify the smoothing priors, define
${H^{(0)}_{\mathit{CA}}}=\frac{1}{\sqrt{n}}\mathbf{1}_n$. Let the county main effect
parameter $\bolds{\Theta}_{\mathit{CO}}$ have prior $\bolds{\Theta}_{\mathit{CO}} \sim
N_{N-1}(\mathbf{0},\tau_0 D^{(-)})$, where $D^{(-)}$ corresponds to
$V^{(-)}$,
that is, $D$ without its $N${th} row and column, $\tau_0>0$ is
unknown and
$\tau_0D^{(-)}$ is a precision matrix. Similarly, let the $j${th}
group of
columns in the cancer-by-county interaction, $V^{(-)}\otimes{{H^{(j)}_{\mathit{CA}}}}$,
have prior $\bolds{\Theta}^{(j)}_{\mathit{CO} \times \mathit{CA}} \sim
N_{N-1}(\mathbf{0},\tau_j D^{(-)})$, for $\tau_j>0$ unknown. Each of
the priors on
$\bolds{\Theta}_{\mathit{CO}}$ and the ${{\bolds{\Theta}^{(j)}_{\mathit{CO} \times \mathit{CA}}}}$
is a CAR prior; the overall level of each CAR, with prior precision
zero, has
been included in the grand mean and cancer main effects.

To compare this to the MCAR model, use SANOVA's priors on $\bolds
{\Theta}$
to produce a marginal prior for $\bolds{\phi}$ comparable to the MCAR's
prior on $\bolds{\phi}$ (Section \ref{ss:mcar}); in other words, integrate
$\bolds{\Theta}_{\mathit{CO}}$ and $\bolds{\Theta}_{\mathit{CO} \times \mathit{CA}}$ out
of the
foregoing setup. A priori,
%
\begin{equation}{\label{precision}}
\left[\matrix{ K &
\pmatrix{
\bolds{\Theta}_{\mathit{CO}} \vspace*{5pt}\cr
\bolds{\Theta}^{(1)}_{{\mathit{CO} \times \mathit{CA}}}\vspace*{3pt}\cr
\vdots\vspace*{5pt}\cr
\bolds{\Theta}^{(n-1)}_{\mathit{CO} \times \mathit{CA}}}}\right]
\end{equation}
has precision $ Q \otimes({H^{(+)}_A}\operatorname{diag}(\tau_j){H^{(+)}_A}{}')$,
where $K$ is the columns of the design matrix for the county main
effects and
cancer-by-county interactions---the right-most $n(N-1)$ columns in equation
(\ref{long})'s design matrix---and ${H^{(+)}_A}=(\frac{1}{\sqrt
{n}}\mathbf{1}_n|H_{\mathit{CA}})$ is an orthogonal matrix. Appendix A in
Zhang, Hodges and Banerjee (\citeyear{Zhang09})
gives a~proof.

\subsubsection{\texorpdfstring{Comparing SANOVA vs MCAR.}{Comparing SANOVA vs MCAR}} \label{ss:comparison}
\begin{figure}[t]

\includegraphics{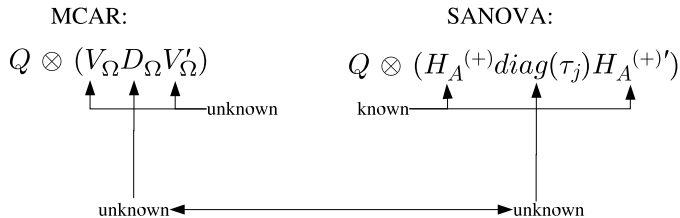}

\caption{Comparing prior precision matrices for $\bolds{\phi}$ in
MCAR and SANOVA.}\label{fig:mcarvssanova}
\end{figure}

Defining $\bolds{\phi}$ as in Sections \ref{ss:mcar} and~\ref{ss:sanovamcar}, consider the MCAR prior for $\bolds{\phi}$, with
within-county precision matrix $\Omega$. Let~$\Omega$ have spectral
decomposition $V_{\Omega}D_{\Omega}V_{\Omega}'$, where $D_{\Omega}$ is
$n \times
n$ diagonal and $V_{\Omega}$ is $n \times n$ orthogonal. Then the prior
precision of $\bolds{\phi}$ is $Q
\otimes(V_{\Omega}D_{\Omega}V_{\Omega}')$, where $Q$ is the known neighbor
relations matrix and $V_{\Omega}$ and $D_{\Omega}$ are unknown.
Comparing MCAR
to SANOVA, the prior precision matrices for the vector $\bolds{\phi
}$ are
as in Figure~\ref{fig:mcarvssanova}.
SANOVA is clearly a special case of MCAR in which $H^{(+)}_A$ is known.
Also, as
described so far, $H^{(+)}_A$ has one column proportional to
$\mathbf{1}_n$
with the other columns being contrasts, while MCAR avoids this
restriction. MCAR
is thus more flexible, while SANOVA is simpler, presumably making it
better identified and easier to set priors for. MCAR should have its
biggest advantage
over SANOVA when the ``true''
$V_{\Omega}$ is not like $H^{(+)}_A$ for any specification of the smoothing
precisions $\tau_j$. However, because data sets often have modest information
about higher-level variances, it may be that using the wrong
$H^{(+)}_A$ usually
has little effect on the analysis. In other words, SANOVA's performance
may be
relatively stable despite having to specify $H^{(+)}_A$, while MCAR may
be more
sensitive to $\Omega$'s prior.

%
%
\subsection[Setting priors in MCAR and SANOVA]{Setting priors in MCAR and SANOVA.}\label{s:prior}
%
\subsubsection{\texorpdfstring{Priors in SANOVA.}{Priors in SANOVA}}\label{ss:prior1}
For the case of normal errors, based on equations~(\ref{datacase}) and
(\ref{long}), setting priors for $\bolds{\Theta},\tau_j,\eta_0$
completes a
Bayesian specification. Since $\bolds{\tau}$ and $\eta_0$ are precision
parameters, one possible prior is Gamma; this paper uses a Gamma with
mean 1 and
variance 10. As mentioned, the grand mean and cancer main effects
$\theta_1,
\theta_2, \theta_3$ have flat priors with $\pi(\theta)\propto1$,
though they
could have proper informative priors. The priors for $\theta_4, \ldots$
are set
according to the SANOVA structure as in Section \ref{ss:sanovamcar}.
We ran
chains drawing in the order~$\theta$, $\tau$ and $\eta_0$ [Appendix B
in Zhang, Hodges and Banerjee (\citeyear{Zhang09}) gives
details]. Hodges, Carlin and Fan (\citeyear{Hodges07}) also considered priors
on the degrees of freedom
in the fitted model, some conditioned so the degrees of freedom in the model's
fit were fixed at a certain degree of smoothness. The present paper emphasizes
comparing MCAR and SANOVA, so we do not consider such priors.
For the case of Poisson errors, we use a normal prior with mean $0$ and variance
$10^6$ for the grand mean and cancer main effects $\theta_1,\theta
_2,\theta_3$.
The other $\theta_i$s are given normal CAR priors as discussed in Section
\ref{ss:sanovamcar}. For the prior on the smoothing precisions $\tau
_j$, we use
Gamma with mean 1 and variance 10. To reduce high posterior
correlations among
the $\theta$s, we used a transformation during MCMC; Appendix C in
Zhang, Hodges and Banerjee~(\citeyear{Zhang09}) gives details.
%
%
\subsubsection{\texorpdfstring{Priors in MCAR.}{Priors in MCAR}}\label{ss:prior2}
MCAR models were fitted in WinBUGS. For the normal-error case, we used
this model
and parameterization:
%
\begin{eqnarray}\label{normal}
Y_{ij} &\sim& N\biggl(\mu_{ij},\frac{1}{\eta_0}\biggr),
\nonumber
\\[-8pt]
\\[-8pt]
\nonumber
\mu_{ij} &=& \beta_j+S_{ij},
\end{eqnarray}
$i=1,\ldots, N; j =1,\ldots,n$, where $\eta_0$ has a gamma prior with
mean 1 and
variance~10 as for SANOVA. To satisfy WinBUGS's constraint that
$\sum_{i}S_{ij}=0$, we add cancer-specific intercepts $\beta_j$. We give
$\beta_j$ a flat prior and for $S$, the spatial random effects, we use an
intrinsic multivariate CAR prior. Similarly, in the Poisson case
%
\begin{eqnarray}\label{poisson}
Y_{ij} &\sim& \operatorname{Poisson}(\mu_{ij}),
\nonumber
\\[-8pt]
\\[-8pt]
\nonumber
\operatorname{log}(\mu_{ij}) &=& \operatorname{log}(E_{ij})+\beta_j+S_{ij},
\end{eqnarray}
where $E_{ij}$ is an offset. Prior settings for $\beta_j$ and $S_{ij}$
are as in
the normal case.
For MCAR priors, the within-county precision matrix $\Omega$ needs a
prior; a Wishart distribution is an obvious choice. If $\Omega\sim
\operatorname{Wishart}(R,\nu)$, then $E(\Omega) = \nu R^{-1}$. We want a ``vague''
Wishart prior; usually $\nu= n$ is used but little is known about how
to specify $R$. Thus, we considered three different $R$s, each
proportional to the identity matrix. One of these priors sets $R$'s
diagonal entries to $R_{ii} = 0.002$, close to the setting used in an
example in the GeoBUGS manual (oral cavity cancer and lung cancer in
West Yorkshire). The other two $R$s are the identity matrix and 200
times the identity. For the special case $n=1$, where the Wishart
reduces to a Gamma, these Wisharts are $\Gamma(0.5,0.001)$, $\Gamma
(0.5,0.5)$ and $\Gamma(0.5,100)$, respectively.
%
%

\section{\texorpdfstring{Simulation experiment.}{Simulation experiment}}\label{S:simulation1}
For this simulation experiment, artificial data were simulated from the
model used in SANOVA with a spatial factor, as described in Section
\ref{ss:sanovamcar}. Three different types of Bayesian analysis were
applied to
the simulated data: SANOVA with the same ${H^{(+)}_A}$ used to generate the\vspace*{-1pt}
simulated data (called ``SANOVA correct''); SANOVA with incorrect ${H^{(+)}_A}$;
and MCAR. SANOVA correct is a theoretical best possible analysis in
that it
takes as known things that MCAR estimates, that is, it uses additional correct
information. SANOVA correct cannot be used in practice, of course,
because the
true ${H^{(+)}_A}$ is\vspace*{-1pt} not known. MCAR vs SANOVA with incorrect
${H^{(+)}_A}$ is
the comparison relevant to practice, and comparing them to SANOVA
correct shows
how much each method pays for its ``deficiency'' relative to SANOVA
correct. Obviously it is not enough to test the SANOVA model using only
data generated from a similar SANOVA model. To avoid needless computing
and facilitate comparisons, instead of generating data from an MCAR
model and fitting a SANOVA model as specified above, we use a trick
that is equivalent to this. Section \ref{six} gives the details.
%
%
\begin{table}[b]
\caption{Experimental conditions in the simulation experiments}\label
{t:dataset}
\begin{tabular*}{\textwidth}{@{\extracolsep{\fill}}lccc@{}}
\hline
{\textbf{Error distribution}}&{$\bolds{\eta}_\mathbf{0}$}&
{$\bolds{(}\bolds{\tau}_\mathbf{0}\bolds{/}\bolds{\eta}_\mathbf{0}\bolds{,}\bolds{\tau
}_\mathbf{1}\bolds{/}\bolds{\eta}_\mathbf{0}\bolds{,}\bolds{\tau}_\mathbf
{2}\bolds{/}\bolds{\eta}
_\mathbf{0}\bolds{)}\bolds{|}\bolds{(}\bolds{\tau}_\mathbf{0}\bolds{,}\bolds{\tau}_\mathbf
{1}\bolds{,}\bolds{\tau}_\mathbf{2}\bolds{)}$}&{\textbf{Data name}}\\
\hline
Normal & 1 & (100, 100, 0.1) & Data1\\
& 1 & (0.1, 100, 0.1) & Data2\\
& 10 & (100, 100, 0.1) & Data3\\
& 10 & (0.1, 100, 0.1) & Data4\\[3pt]
$\operatorname{Poisson}$ & NA & (100, 100, 0.1) & Data5\\
& NA & (0.1, 100, 0.1) & Data6\\
\hline
\end{tabular*}
\end{table}

\subsection{\texorpdfstring{Design of the simulation experiment.}{Design of the simulation experiment}}\label{Design}
We simulated both normally-distributed and Poisson-distributed data.
For both
types of data, we considered two different true sets of smoothing parameters
$\mathbf{r} =\bolds{\tau}/\eta_0$ or $\bolds{\tau}$ (Table
\ref{t:dataset}).
For the normal data, we considered $\bolds{\tau}/\eta_0$, since this ratio
determines smoothing in normal models, and we also considered two error
precisions $\eta_0$ (Table \ref{t:dataset}).
%
%
\subsubsection{\texorpdfstring{Generating the simulated data sets.}{Generating the simulated data sets}} \label{S:generatedata}
To generate data from the\break SANOVA model, we need to define the true
${H^{(+)}_A}$. Let
\[
{\mathit{HA}}_1 = \pmatrix{
1 & -2 & 0\cr
1 & 1 & -1\cr
1 & 1 & 1}
\pmatrix{
\frac{1}{\sqrt{3}} & 0 & 0\cr
0 & \frac{1}{\sqrt{6}} & 0 \cr
0 & 0 & \frac{1}{\sqrt{2}}\cr}
.
\]
We used ${\mathit{HA}}_1$ as the correct ${H^{(+)}_A}$; its columns are scaled
to have
length 1. Given $V^{(-)}$ and with ${H^{(+)}_A}$ known, one draw of
$\bolds{\Theta}$ and $\epsilon$ produces a draw of $X_D\bolds{\Theta
}+\epsilon$,
therefore a
draw of $\mathbf{y}$. In the simulation, we let the grand mean and
main effects,
which are not smoothed, have true value 5. Each observation is
simulated from a
$3 \times20$ factorial design, where 3 is the number of cancers and 20
is the
number of counties. We used the 20 counties in the right lower corner of
Minnesota's map, with their actual neighbor relations. Thus, the
dimension of
each artificial data set is 60.
The simulation experiment is a repeated-measures design, in which a ``subject''
$s$ in the design is a draw of $(\bolds{\delta}^{(s)},\bolds{\gamma}^{(s)})$, referring to equation (\ref{dataconstraint}), where
$\bolds{\delta}^{(s)}_{1-3}=5$ and ${\bolds{\delta}^{(s)}_{4-60}} \sim
N_{57}(\mathbf{0},I_3 \otimes D^{(-)})$ specify
$\bolds{\Theta}$ and $\bolds{\gamma}^{(s)} \sim N_{60}(\mathbf
{0},I_{60})$ gives $\bolds{\epsilon}$. For the
normal-errors case, 100 such ``subjects'' were generated. Given a
design cell in
the simulation experiment with $\bolds{\tau}=(a,b,c)$ and $\eta
_0=d$, the
artificial data set for subject\vspace*{-1pt} $s$ is $\mathbf{y}^{(s)} = X_D\textrm{
}\operatorname{diag}(\mathbf{1}'_3,\frac{1}{\sqrt{a}}\mathbf{1}'_{19},$ $\frac
{1}{\sqrt
{b}}\mathbf{1}'_{19},\frac{1}{\sqrt{c}}\mathbf{1}'_{19})\bolds
{\delta
}^{(s)}+\frac{1}{\sqrt
{d}}\bolds{\gamma}^{(s)}$. All factors of the simulation experiment
were applied to
each of
the 100 ``subjects.'' For the normally-distributed data, the simulation
experiment had these factors: (a) the true
$(\tau_0/\eta_0,\tau_1/\eta_0,\tau_2/\eta_0)\dvt (100,100,0.1)$ or
$(0.1,100,0.1)$;
(b) the true error precision $\eta_0$: 1 or 10; and (c) six
statistical methods,
described below in Section \ref{six}. Each design cell described in Table~\ref{t:dataset} thus had 100 simulated data sets.
Similarly, for the Poisson-data experiment, another 100 ``subjects'' were
generated, but now there is no $\bolds{\gamma}^{(s)}$. Thus, each
``subject'' $s$ is
a vector $\bolds{\delta}^{(s)}$, where $\bolds{\delta}^{(s)}$ is as
described
above. For
the design cell with $\bolds{\tau}=(100,100,0.1)$, the artificial data
for subject
$s$ is $\mathbf{y}^{(s)} \sim \operatorname{Poisson}(\bolds{\mu}^{(s)})$, where
$\operatorname{log}(\bolds{\mu}^{(s)})=
\operatorname{log}(\mathbf{E}) + X_D\operatorname{diag}(\mathbf{1}'_3,\frac
{1}{10}\mathbf
{1}'_{19},\frac{1}{10}\mathbf{1}'_{19},\frac{1}{\sqrt{0.1}}\mathbf
{1}'_{19})\bolds{\delta}^{(s)}$. In the simulation experiment, we use
``internal standardization''
of the Minnesota 3-cancer data to supply the expected numbers of cancers
$E_{ij}$. Among the 20 extracted counties, Hennepin county has the largest
average population over 11 years, about 1.1 million; its cancer counts are
5294, 119 and 439 for lung, larynx and esophagus respectively.
Faribault county
has the smallest average population, 16,501, with cancer counts 110, 7
and 13
respectively. The $E_{ij}$ have ranges 80 to 5275, 2 to 113 and 7 to
449 for
lung, larynx and esophagus cancer respectively. For the Poisson data, the
simulation experiment had these factors: (a) the true
$\tau_0,\tau_1,\tau_2$: $(100,100,0.1)$ or $(0.1,100,0.1)$; and (b) six
statistical
methods described below in Section \ref{six}. Again, each of the two design
cells in Table \ref{t:dataset} had 100 simulated data sets.
%
%
\begin{table}[b]
\caption{The six statistical methods used in the simulation
experiment}\label{t:procedure}
\begin{tabular*}{265pt}{@{\extracolsep{\fill}}lcc@{}}
\hline
{\textbf{Procedure}}&\textbf{Prior}\\
\hline
SANOVA with correct $H_A^{(+)}$ & $ \eta_0,\tau_j \sim\Gamma
(0.1,0.1)$ for
$j=0,1,2$ \\
SANOVA with incorrect $H_A^{(+)}$ & $\eta_0,\tau_j \sim\Gamma
(0.1,0.1)$ for
$j=0,1,2$ \\
Variant SANOVA with $H_{\mathit{AM}}$ & $\eta_0,\tau_j \sim\Gamma(0.1,0.1)$ for
$j=0,1,2$ \\
MCAR & $\Omega\sim\operatorname{Wishart}(R,3)$, $R=0.002I_3$ \\
MCAR & $\Omega\sim\operatorname{Wishart}(R,3)$, $R=I_3$ \\
MCAR & $\Omega\sim\operatorname{Wishart}(R,3)$, $R=200I_3$\\
\hline
\end{tabular*}
\end{table}

\subsubsection{\texorpdfstring{The six methods (procedures).}{The six methods (procedures)}}\label{six}
For each simulated data set, we did a~Bayesian analysis for each of six
different
models described in Table \ref{t:procedure}. The six models are:
SANOVA with
the correct ${H^{(+)}_A}$, $\mathit{HA}_1$; SANOVA with a somewhat incorrect
${H^{(+)}_A}$, $\mathit{HA}_2$ given below; a variant SANOVA with a very incorrect
${H^{(+)}_A}$, $H_{\mathit{AM}}$ given below; MCAR with $R_{ii} = 0.002$; MCAR with
$R_{ii} = 1$; and MCAR with $R_{ii} = 200$ (see Section \ref{ss:prior2}). $\mathit{HA}_2$
and $H_{\mathit{AM}}$ are
\begin{eqnarray*}
{\mathit{HA}}_2 &=& \pmatrix{
1 & 1 & 1\cr
1 & -2 & 0\cr
1 & 1 & -1}
\pmatrix{
\frac{1}{\sqrt{3}} & 0 & 0\cr
0 & \frac{1}{\sqrt{6}} & 0 \cr
0 & 0 & \frac{1}{\sqrt{2}}},\\
H_{\mathit{AM}} &=& \pmatrix{
0.56 & -0.64 & -0.52\cr
-0.53 & -0.77 & 0.36\cr
-0.63 & 0.07 & -0.77}.
\end{eqnarray*}
The incorrect $\mathit{HA}_2$ has the same first column (grand mean)
as the correct
$\mathit{HA}_1$, so it differs from the correct $\mathit{HA}_1$, though less than it
might. As noted above, we need to see how the SANOVA model performs for
data generated from an MCAR model in which $V_\Omega$ from Figure \ref
{fig:mcarvssanova} does not have a column proportional to $\mathbf
{1}_n$. To
do this without needless computing, we used a trick: we used the data
generated from a~SANOVA model with $\mathit{HA}_1$ and fit the variant SANOVA
mentioned above, in which $H_A^{(+)}$ is replaced by the orthogonal
matrix $H_{\mathit{AM}}$ with no column proportional to $\mathbf{1}_n$, chosen
to be
very different from $\mathit{HA}_1$. For normal errors (Data1 through Data4),
this is precisely equivalent to fitting a SANOVA with $H_A^{(+)} =
\mathit{HA}_1$ to data generated from an MCAR model with $V_\Omega= B \mathit{HA}_1$,
for $B = \mathit{HA}_1 H_{\mathit{AM}}^{-1}$, that is,
%
\begin{equation}
V_{\Omega} =
\left[\matrix{
0.43 & -0.74 & -0.52 \cr
-0.13 & -0.63 & 0.77 \cr
-0.89 & -0.26 & -0.37}\right]
\end{equation}
(to 2 decimal places). For Poisson errors (Data5, Data6), the
equivalence is no longer precise but the divergence of fitted SANOVA
[using ${H^{(+)}_A} = H_{\mathit{AM}}$] and generated data [using ${H^{(+)}_A} =
\mathit{HA}_1$] is quite similar.
Finally, we considered three priors for MCAR because little is known
about how to set this prior and we did not want to hobble MCAR with an
ill-chosen prior. For the SANOVA and variant SANOVA analyses, we gave
$\tau_j$ a $\Gamma(0.1,0.1)$ prior with mean 1 and variance 10 for both
the normal data and the Poisson data.
%
%
\subsection{\texorpdfstring{Outcome measures.}{Outcome measures}}
To compare the six different methods for normal and Poisson data, we consider
three criteria. The first is average mean squared error (AMSE). For
each of the
60 ${(X_D\bolds{\Theta})}_{ij}$, the mean squared error is defined as
the average
squared error over the 100 simulated data sets. AMSE for each design
cell in the
simulation experiment is defined as the average of mean squared error
over the
60 ${(X_D\bolds{\Theta})}_{ij}$. Thus, for the design cell labeled
DataK in Table
\ref{t:dataset}, define
%
\begin{equation}
\widehat{\mathit{AMSE}}_K=
\frac{1}{L}\sum^L_{d=1}\sum^{N}_{i=1}\sum^n_{j=1}[(X_D\widehat
{\bolds
{\Theta}})^d_{ij}-{(X_D\bolds{\Theta})}^d_{ij}]^2/Nn,
\end{equation}
where $ L=100, N=20,n=3, K=1,\ldots,4$ for Normal, $K=5,6$ for Poisson,
$\bolds{\Theta}$ is the true value and $\widehat{\bolds{\Theta}}$ is
the posterior
median of
$\bolds{\Theta}$. For each design cell~($K$), the Monte Carlo standard
error for
AMSE is $(100)^{-0.5}$ times the standard deviation, across DataK's 100
simulated data sets, of $\sum^{N}_{i=1}\sum^n_{j=1}[(X_D\widehat
{\bolds
{\Theta}})^d_{ij}-{(X_D\bolds{\Theta})}^d_{ij}]^2/Nn$.
The second criterion is the bias of $X_D\bolds{\Theta}$. For each of
DataK's 100
simulated data sets, first compute posterior medians of ${(X_D\bolds
{\Theta})}_{1,1},\ldots,{(X_D\bolds{\Theta})}_{20,3}$, then average
each of
those posterior
medians across the 100 simulated data sets. From this average, subtract
the true
${(X_D\bolds{\Theta})}_{ij}$s to give the estimated bias for each of
the 60
${(X_D\bolds{\Theta})}_{ij}$s. MBIAS is defined as the
2.5{th}, 50{th} and
$97.5${th} percentiles of the 60 estimated biases. More explicitly,
for design
cell DataK, MBIAS is
%
\begin{eqnarray}
\widehat{\mathit{MBIAS}}_K &=& 2.5{\mathrm{th}},50{\mathrm{th}},97.5{\mathrm{th}}\mbox{ percentiles of
}
\nonumber
\\[-8pt]
\\[-8pt]
\nonumber
&&\Biggl(
\frac{1}{L}\sum^L_{d=1}(X_D\widehat{\bolds{\Theta}}^d-{X_D\bolds
{\Theta}
}^d)\Biggr).
\end{eqnarray}
Finally, the coverage rate of Bayesian $95\%$ equal-tailed posterior intervals,
``PI rate,'' is the average coverage rate for the 60 individual ${(X_D
\bolds{\Theta})}_{ij}$s.

\subsection{\texorpdfstring{Markov chain Monte Carlo specifics.}{Markov chain Monte Carlo
specifics}}
While the MCAR models were implemented in WinBUGS, our SANOVA
implementations were coded in R and run on Unix. The different
architectures do not permit a fair comparison between the run times of
SANOVA and MCAR. However, the SANOVA models have lower computational
complexity than the MCAR models: MCAR demands a spectral decomposition
in every iteration, while SANOVA does not. For each of our models, we
ran three parallel MCMC chains for 10,000 iterations. The CODA package
in R (\href{http://www.r-project.org}{www.r-project.org}) was used to diagnose
convergence by monitoring mixing using Gelman--Rubin diagnostics and
autocorrelations [e.g., Gelman et al. (\citeyear{Geetal2004}), Section
11.6]. Sufficient
mixing was seen within 500 iterations for the SANOVA models, while 200
iterations typically revealed the same for the MCAR models; we retained
$8000 \times3$ samples for the posterior analysis.
%

\begin{table}
\caption{AMSE for simulated normal and Poisson data } \label{t:AMSE}
\begin{tabular*}{\textwidth}{@{\extracolsep{\fill}}lcccccccc@{}}
\hline
&\multicolumn{4}{c}{\textbf{Normal-error model}}& &\multicolumn{3}{c@{}}{\textbf{Poission-error model}} \\[-6pt]
&\multicolumn{4}{c}{\hrulefill}& &\multicolumn{3}{c@{}}{\hrulefill} \\
\textbf{Model}&\textbf{Data1}&\textbf{Data2}&\textbf{Data3}&\textbf{Data4}&&\textbf{Data5}&&\textbf{Data6}\\
\hline
SANOVA with $\mathit{HA}_1$ & 0.34&0.60&0.04&0.06& &0.02&\mbox{\quad   }&0.04\\
SANOVA with $\mathit{HA}_2$ & 0.47&0.84&0.05&0.07& &0.02&&0.14\\
SANOVA with $\mathit{HA}_{\mathit{AM}}$ & 0.48&0.74&0.05&0.06& &0.03&&0.11\\
MCAR with $R_{ii}=0.002$ &0.66&1.88&0.04&0.13& &0.02&&0.04\\
MCAR with $R_{ii}=1$ &0.36&0.84&0.04&0.06& &0.02&&0.06\\
MCAR with $R_{ii}=200$ &0.93&0.92&0.09&0.09& &0.24&&0.36\\
\hline
\end{tabular*}
\end{table}

\begin{figure}[b]

\includegraphics{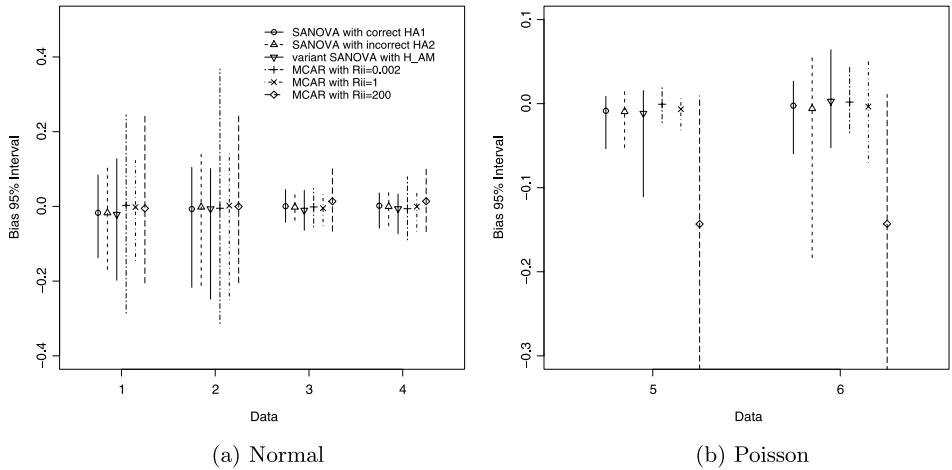}

\caption{MBIAS for simulated normal and Poisson data.}\label{fig:BIAS1}
\end{figure}

%
\subsection{\texorpdfstring{Results.}{Results}}
\begin{figure}

\includegraphics{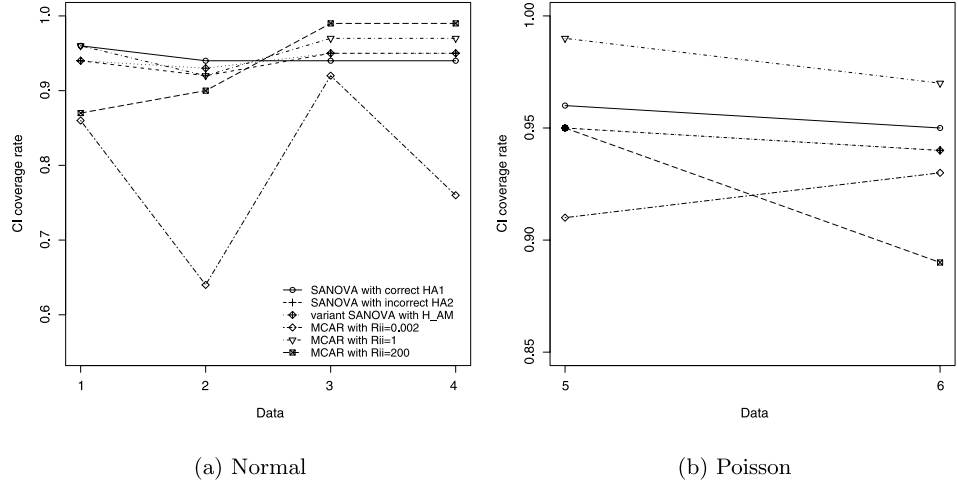}

\caption{PI rate for simulated normal and Poisson data.}\label{fig:CI1}
\end{figure}

Table \ref{t:AMSE} and Figures \ref{fig:BIAS1} and \ref{fig:CI1} show
the simulation experiment's results. Table \ref{t:AMSE} shows AMSE; for
all methods and design cells, the standard Monte Carlo errors of AMSE
are small, less than 0.07, 0.005 and 0.025 for Data1$/$Data2,
Data3$/$Data4 and Data5$/$Data6 respectively. Figure \ref{fig:BIAS1}
shows MBIAS, where the middle symbols represent the median bias and the
line segments represent the $2.5${th} and $97.5${th} percentiles.
Figure \ref{fig:CI1} shows coverage of the $95\%$ posterior intervals.
Denote SANOVA with the correct ${H^{(+)}_A}$ ($\mathit{HA}_1$) as ``SANOVA
correct,'' SANOVA with $\mathit{HA}_2$ as ``SANOVA incorrect,'' the variant
SANOVA with $H_{\mathit{AM}}$ as ``SANOVA variant,'' MCAR with $R_{ii}=0.002$ as
``MCAR0.002'' and so on.

%
%
\subsubsection{\texorpdfstring{As expected, SANOVA with correct ${H^{(+)}_A}$ performs
best.}{As expected, SANOVA with correct ${H^{(+)}_A}$ performs
best}}
For normal data, SANOVA correct has the smallest AMSE for all true
$\eta
_0$ and $\bolds{\tau}$ (Table \ref{t:AMSE}). The advantage is larger in
Data1 and Data2 where the error precision $\eta_0$ is 1 than in Data3
and Data4 where $\eta_0$ is 10 (i.e., error variation is smaller). For
Poisson data, SANOVA correct also has the smallest AMSE. Considering
MBIAS (Figure~\ref{fig:BIAS1}), SANOVA correct has the narrowest MBIAS
intervals for all cases. In Figure \ref{fig:CI1}, the posterior
coverage for SANOVA correct is nearly nominal. As expected, then,
SANOVA correct performs best among the six methods.
%
%
\subsubsection{\texorpdfstring{SANOVA with incorrect $\mathit{HA}_2$ and $H_{\mathit{AM}}$ perform very well.}
{SANOVA with incorrect $\mathit{HA}_2$ and $H_{\mathit{AM}}$ perform very
well}}
Table \ref{t:AMSE} shows that, for normal data, both SANOVA incorrect
and SANOVA variant have smaller AMSEs than MCAR200 and MCAR0.002, and
AMSEs at worst close to MCAR1's. For Poisson data, Table \ref{t:AMSE}
shows that MCAR0.002 and MCAR1 do somewhat better than SANOVA incorrect
and variant SANOVA.
Considering MBIAS in normal data [Figure \ref{fig:BIAS1}(a)], the width
of the
$95\%$ MBIAS intervals for SANOVA incorrect are the same as or smaller
than for
all three MCAR procedures. Similarly, SANOVA variant has MBIAS
intervals better
than MCAR0.002 and MCAR200 and almost as good as MCAR1. Figure
\ref{fig:BIAS1}(b)
for Poisson data shows SANOVA correct, MCAR0.002 and MCAR1 have similar MBIAS
intervals. SANOVA variant in Data5 and SANOVA incorrect in Data6 show
the worst
performance for MBIAS apart from MCAR200, whose MBIAS interval is much the
widest.
Figure \ref{fig:CI1}(a) shows that, for normal data, interval coverage
for SANOVA
incorrect and SANOVA variant is very close to nominal. It appears that the
specific value of ${H^{(+)}_A}$ has little effect on PI coverage rate
for the cases considered here. Apart
from MCAR200 for Data1$/$Data2 and MCAR0.002 for Data1 through Data4,
which show
low coverage, all the other methods have coverage rates greater than
$90\%$ for
normal data, most close to $95\%$. For Data3 and Data4, PI rates for MCAR200
reach above $99\%$. For Poisson data, the PI rates for SANOVA incorrect and
SANOVA variant are close to nominal and better than MCAR0.002 and
MCAR200. In
particular, all SANOVAs have the closest to nominal coverage rates for both
normal and Poisson data, which again shows the stability of SANOVA under
different ${H^{(+)}_A}$ settings.
%
%
\subsubsection{\texorpdfstring{MCAR is sensitive to the prior on $\Omega$.}{MCAR is sensitive to the prior on
$\Omega$}}
To fairly compare SANOVA and MCAR, we considered MCAR under three different
prior settings. For normal data, MCAR1 has the smallest AMSEs and narrowest
MBIAS intervals among the MCARs considered, while MCAR0.002 has the
largest and
widest, respectively. For Poisson data, however, MCAR0.002 has the best
AMSE and
MBIAS among the MCARs. MCAR200 performs poorly for both Normal and
Poisson. The
coverage rates in Figure \ref{fig:CI1} show similar comparisons. These results
imply that the prior matters for MCAR: no single prior was always best. By
comparison, SANOVA seems more robust, at least for the cases considered.
%
%
\subsection{\texorpdfstring{Summary.}{Summary}}
As expected, SANOVA correct had the best performance because it uses more
correct information. For normal data, SANOVA incorrect and SANOVA
variant had
similar AMSEs, better than two of the three MCARs for the data sets considered.
For Poisson data, SANOVA incorrect and SANOVA variant had AMSEs as good
as those
of MCAR0.002 and MCAR1 for Data5 and somewhat worse for Data6, while
showing nearly nominal coverage rates in all cases
and less tendency to bias than MCAR in most cases. Replacing the
$\Gamma(0.1,0.1)$ prior for $\bolds{\tau}$ with $\Gamma
(0.001,0.001)$ left
AMSE and
MBIAS almost unchanged and coverage rates a bit worse (data not shown). MCAR,
on the other hand, seems more sensitive to the prior on $\Omega$.
MCAR0.002 tends to smooth more than MCAR1, more so in normal models
where the prior is more influential than in Poisson models. (The latter
is true because data give more information about means than variances,
and the Poisson model's error variance is the same as its mean, while
the normal model's is not.) For the normal data, MCAR0.002's tendency
to extra shrinkage appears to make it oversmooth and perform poorly for
Data2 and Data4, where the truth is least smooth. For the Poisson data,
MCAR0.002 and MCAR1 give results similar to each other and somewhat
better than the SANOVAs except for interval coverage.
Therefore, SANOVA, with stable results under different ${H^{(+)}_A}$
and with parameters that are easier to understand and interpret, may be
a good competitor to MCAR in multivariate spatial
smoothing.
%
%
\section{\texorpdfstring{Example: Minnesota 3-cancer data.}{Example: Minnesota 3-cancer data}}\label{sec4}
\begin{figure}

\includegraphics{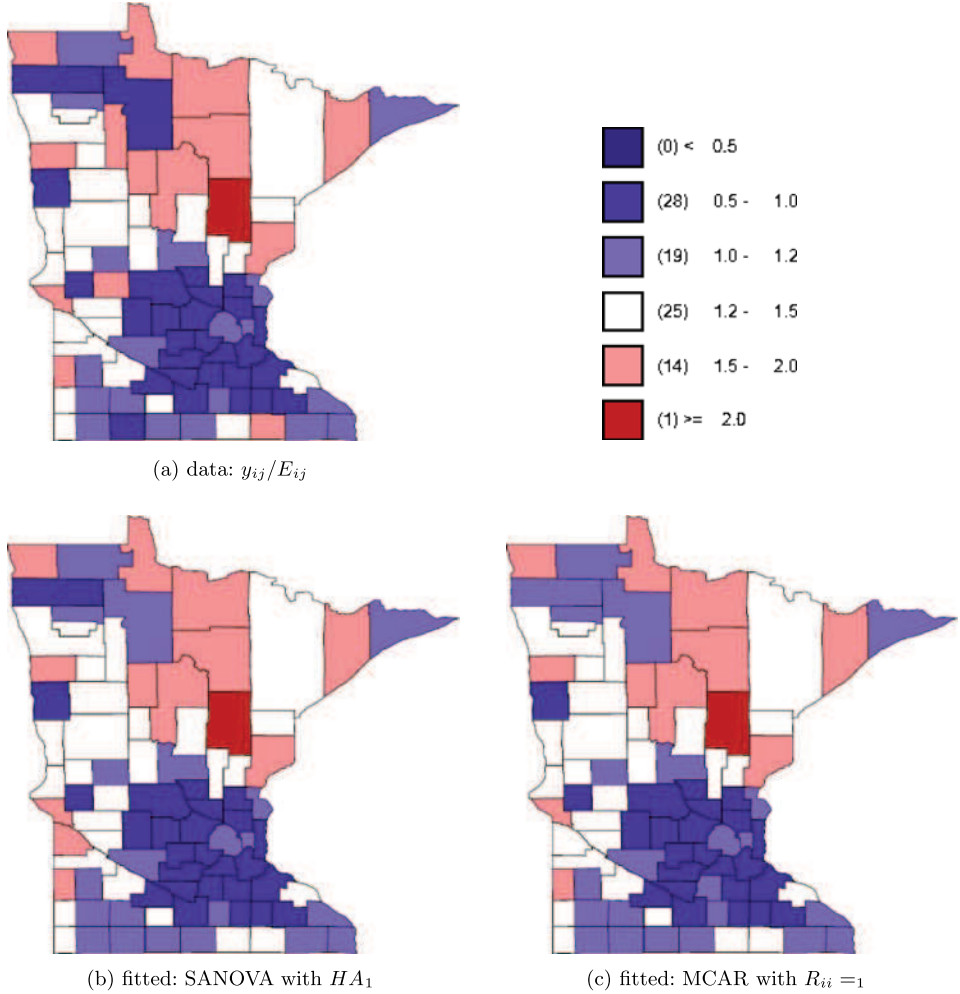}

\caption{Lung cancer data and fitted values.} \label{map1}
\end{figure}

\begin{figure}

\includegraphics{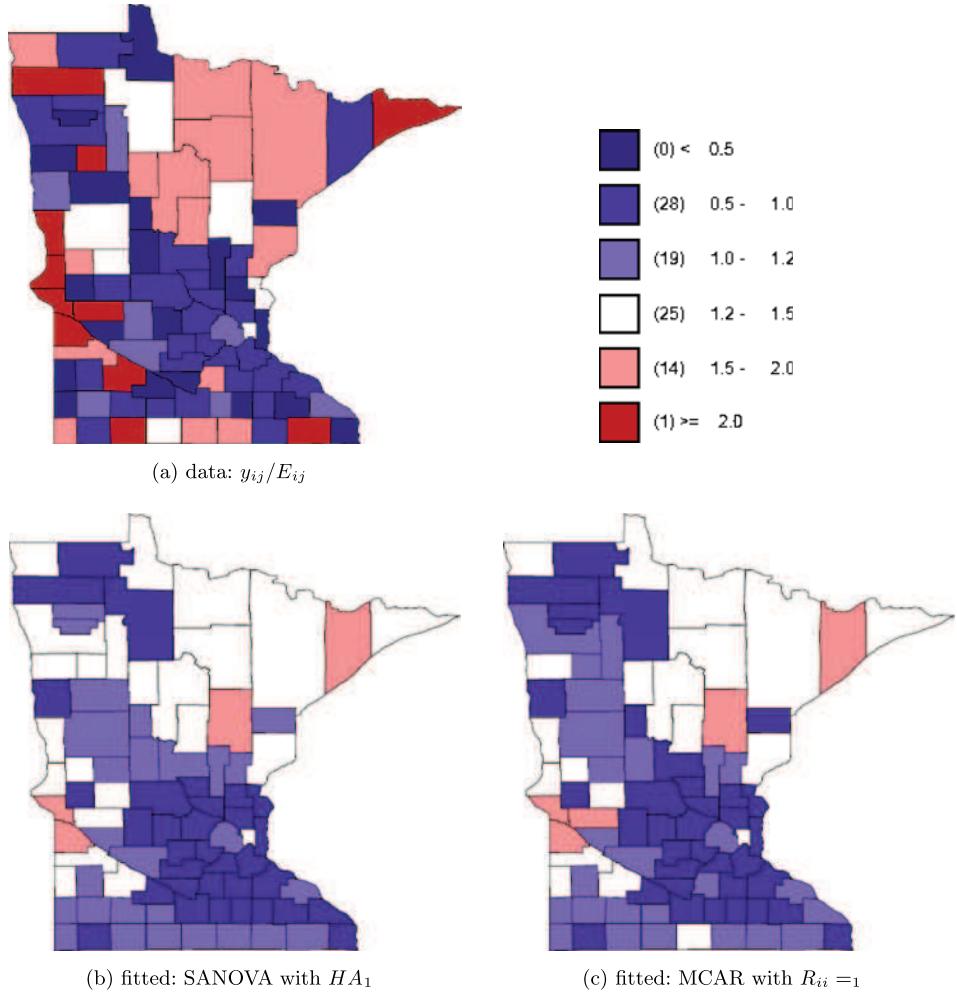}

\caption{Larynx cancer data and fitted values. }\label{map2}
\end{figure}

\begin{figure}

\includegraphics{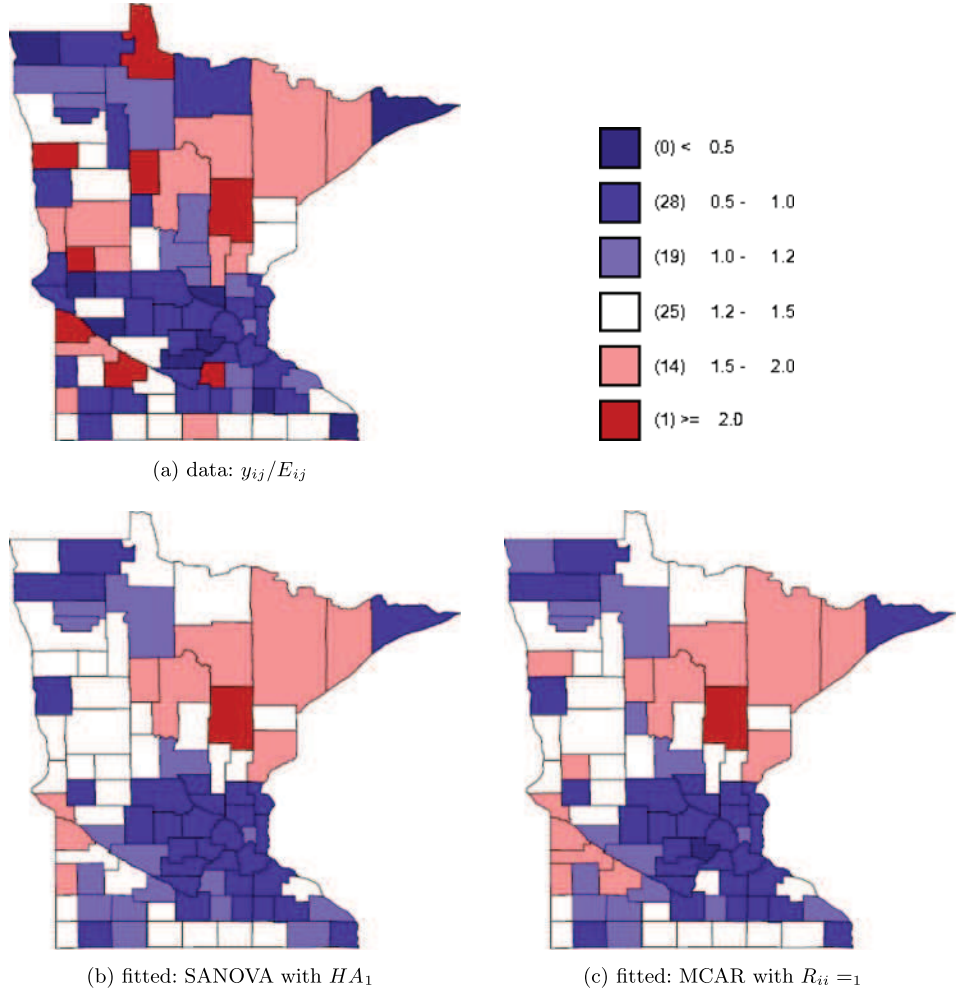}

\caption{Esophagus cancer data and fitted values. }\label{map3}
\end{figure}

\begin{figure}[t]

\includegraphics{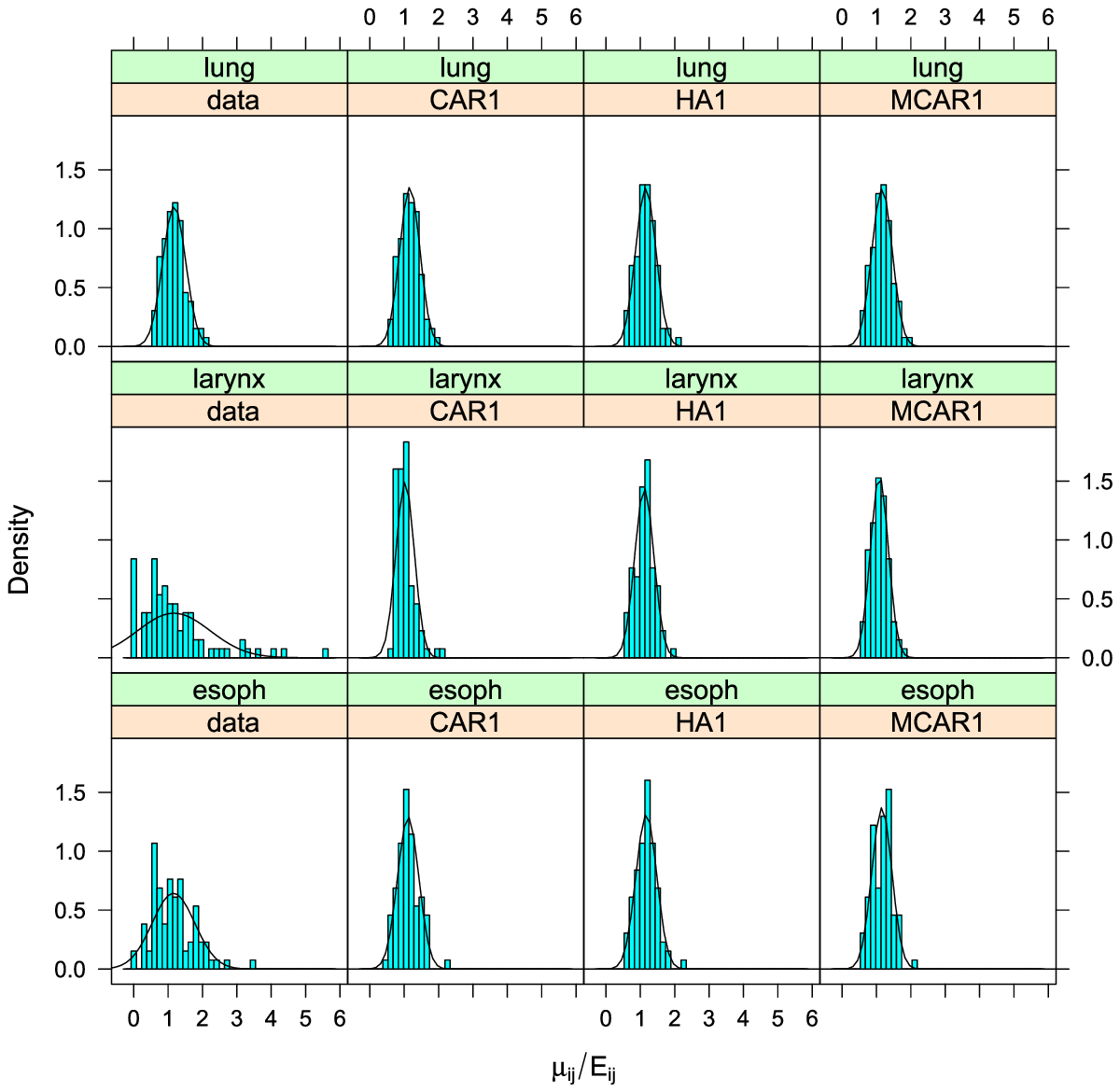}

\caption{ Comparing data and fitted values for each cancer. The
``Data'' panel
shows the density for $y_{ij}/E_{ij}$, while the other three panels
show the posterior median of $\mu_{ij}/E_{ij}$ for univariate CAR,
SANOVA and MCAR.}\label{t:counts}
\end{figure}

Researchers in different fields have illustrated that accounting for
spatial correlation could provide insights that would have been
overlooked otherwise, while failure to account for spatial association
could potentially lead to spurious and sometimes misleading results
[see, e.g., Turechek and Madden (\citeyear{Turechek02}),
Ramsay, Burnett and Krewski (\citeyear{Ramsay03}),
Lichstein et al. (\citeyear{Lichstein02})]. Among the widely
investigated diseases are
the different types of cancers.
We applied SANOVA and MCAR to a cancer-surveillance data set describing
total incidence counts of 3 cancers (lung, larynx, esophagus) in
Minnesota's 87 counties for the years 1990 to 2000 inclusive.
Minnesota's geography and history make it plausible that disease
incidence would show spatial association. Three major North American
land forms meet in Minnesota: the Canadian Shield to the north, the
Great Plains to the west, and the eastern mixed forest to the
southeast. Each of these regions is distinctive in both its terrain and
its predominant economic activity: mining and outdoors tourism in the
mountainous north, highly mechanized crop cultivation in the west, and
dairy farming in the southeast. The different regions were also settled
by somewhat different groups of in-migrants, for example,
disproportionately many Scandinavians in the north. These factors imply
spatial association in occupational hazards as well as culture,
weather, and access to health care especially in the thinly-populated
north, which might be expected to produce spatial association in diseases.
With multiple cancers one obvious option is to fit a separate
univariate model for each cancer. But diseases may share the same
spatially distributed risk factors, or the presence of one disease
might encourage or inhibit the presence of another in a region, for
example, larynx and esophagus cancer have been shown to be closely
related spatially [Baron et al.~(\citeyear{Baron93})]. Thus, we may
need to account
for dependence among the different cancers while maintaining spatial
dependence between sites.
Although the data set has counts broken out by age groups, for the
present purpose we ignore age standardization and just consider total
counts for each cancer. Age standardization would affect only the
expected cancer counts $E_{ij}$, while other covariates could be added
to either SANOVA or MCAR as unsmoothed fixed effects (i.e., in the
$A_1$ design matrix).
Given the population and disease count of each county, we estimated the expected
disease count for each cancer in each county using the Poisson model.
Denote the
$87 \times3$ counts as $y_{1,1},\ldots
,y_{
87,3}$; then the model is
%
\begin{eqnarray}
y_{ij}|\mu_{ij} &\sim& \operatorname{Poisson}(\mu_{ij}),
\nonumber
\\[-8pt]
\\[-8pt]
\nonumber
\operatorname{log}(\mu_{ij})&=& \operatorname{log}(E_{ij})+(X_D\bolds{\Theta})_{ij},
\end{eqnarray}
where $X_D\bolds{\Theta}$ is the SANOVA structure and $\bolds{\Theta
}$ has
priors as in
Section \ref{ss:sanovamcar}. For disease $j$ in county~$i$, $E_{ij}=
P_{i}\frac{\sum_{i}O_{ij}}{\sum_{i}P_{i}}$, where $O_{ij}$ is the
disease count
for county~$i$\vspace*{1pt} and disease $j$ and $P_i$ is county~$i$'s population.
For the
SANOVA design matrix, we consider $\mathit{HA}_1$ and $\mathit{HA}_2$ from the simulation
experiment, though now neither is known to be correct. We also consider a
variant SANOVA analysis using ${H^{(+)}_A}$ estimated from the MCAR1
model, to
test the stability of the SANOVA results. Appendix D in Zhang, Hodges
and Banerjee
(\citeyear{Zhang09}) describes the latter
analysis.
Figures \ref{map1} to \ref{map3} show the data and results for MCAR1
and SANOVA with~$\mathit{HA}_1$. In each figure, the upper left plot shows the
observed $y_{ij}/E_{ij}$; the two lower plots show the posterior median
of $\mu_{ij}/E_{ij}$ for MCAR1 and SANOVA with $\mathit{HA}_1$. Lung cancer
counts tended to be high and thus were not smoothed much by any method,
while counts of the other cancers were much lower and thus smoothed
considerably more (see also Figure \ref{t:counts}). Since SANOVA with
$\mathit{HA}_1$, $\mathit{HA}_2$ and estimated $H_A^{(+)}$ gave very similar results,
only those for $\mathit{HA}_1$ are shown. Results for MCAR0.002 are similar to
those for MCAR1, so they are omitted. As expected, MCAR200 shows the
least shrinkage among the three MCARs and gives some odd $\mu_{ij}/E_{ij}$.
To compare models, we calculated the Deviance Information Criterion [DIC;
Spiegelhalter et al. (\citeyear{Spetal2002})]. To define DIC,
define the deviance
$D(\bolds{\theta})=-2\log f(\mathbf{y}|\bolds{\theta})+2\log
h(\mathbf{y})$,
where $\bolds{\theta}$ is the parameter vector in the likelihood and
$h(\mathbf{y})$ is a function of the data. Since $h$ does not affect model
comparison, we set $\log h(\mathbf{y})$ to 0. Let $\overline{\bolds
{\theta}}$ be
the posterior mean of $\bolds{\theta}$ and $\overline{D}$ the posterior
expectation
of $D(\bolds{\theta})$. Then define $p_D =\overline{D(\bolds{\theta})}-
D(\overline{\bolds{\theta}})$ to be a measure of model complexity
and define
$\mathit{DIC} = \overline{D}+p_D$.
Table \ref{DIC} shows $\overline{D},p_D$ and $\mathit{DIC}$ for nine analyses,
SANOVA with 3 different ${H^{(+)}_A}$, MCAR with 3 different priors
for~$\Omega$, and 3 fits of univariate CAR models to the individual
diseases, discussed below. Considering~$\overline{D}$, the three
SANOVAs and MCAR1 are similar; \mbox{Figures} \ref{map1} to~\ref{map3} show
the fits are indeed similar. Figure \ref{t:counts} reinforces this
point, showing that MCAR1 and SANOVA with $\mathit{HA}_1$ induce similar
smoothing for the three cancers. SANOVA with $H_A^{(+)}$ estimated from
MCAR has the smallest $\overline{D}$ ($1458$), though its model
complexity penalty ($p_D=103$) is higher than MCAR0.002's ($p_D=79$).
Despite having the second worst fit ($\overline D$), MCAR0.002 has the
best DIC,
and the three SANOVAs have DICs much closer to MCAR0.002's than to the other
MCARs. Generally, all SANOVA models have similar $\overline{D}$
($\approx
1460$) and $\mathit{DIC}$ ($\approx1562$), while MCAR results are sensitive to
$\Omega$'s prior, consistent with the simulation experiment.
For comparison, we fit separate univariate CAR models to the three
diseases considering three different priors for the smoothing
precision, $\tau\sim \operatorname{Gamma}(a,a)$ for $a = 0.001$, $1$ and $1000$. For
each prior, we added up $\overline{D}$, $p_D$ and $\mathit{DIC}$ for three
diseases (see Table \ref{DIC}). With $a=0.001$ and $1$, we obtained
$\overline{D}$'s (1461 and 1453 respectively) competitive with SANOVA,
MCAR0.002 and MCAR1 but with considerably greater complexity penalties
(141 and 149 respectively) and thus DICs slightly larger than 1600. For
$a=1000$, we obtained an even lower $\overline{D}$ $(1432),$ but an
increased penalty (180) resulted in a poorer $\mathit{DIC}$ score. Figure \ref
{t:counts} shows fitted values for CAR1, which were smoothed like MCAR1
and SANOVA for lung and esophagus cancers but smoothed rather more for
larynx cancer. Overall, these results reflect some gain in performance
from accounting for the space-cancer interactions/associations.

\begin{figure}

\includegraphics{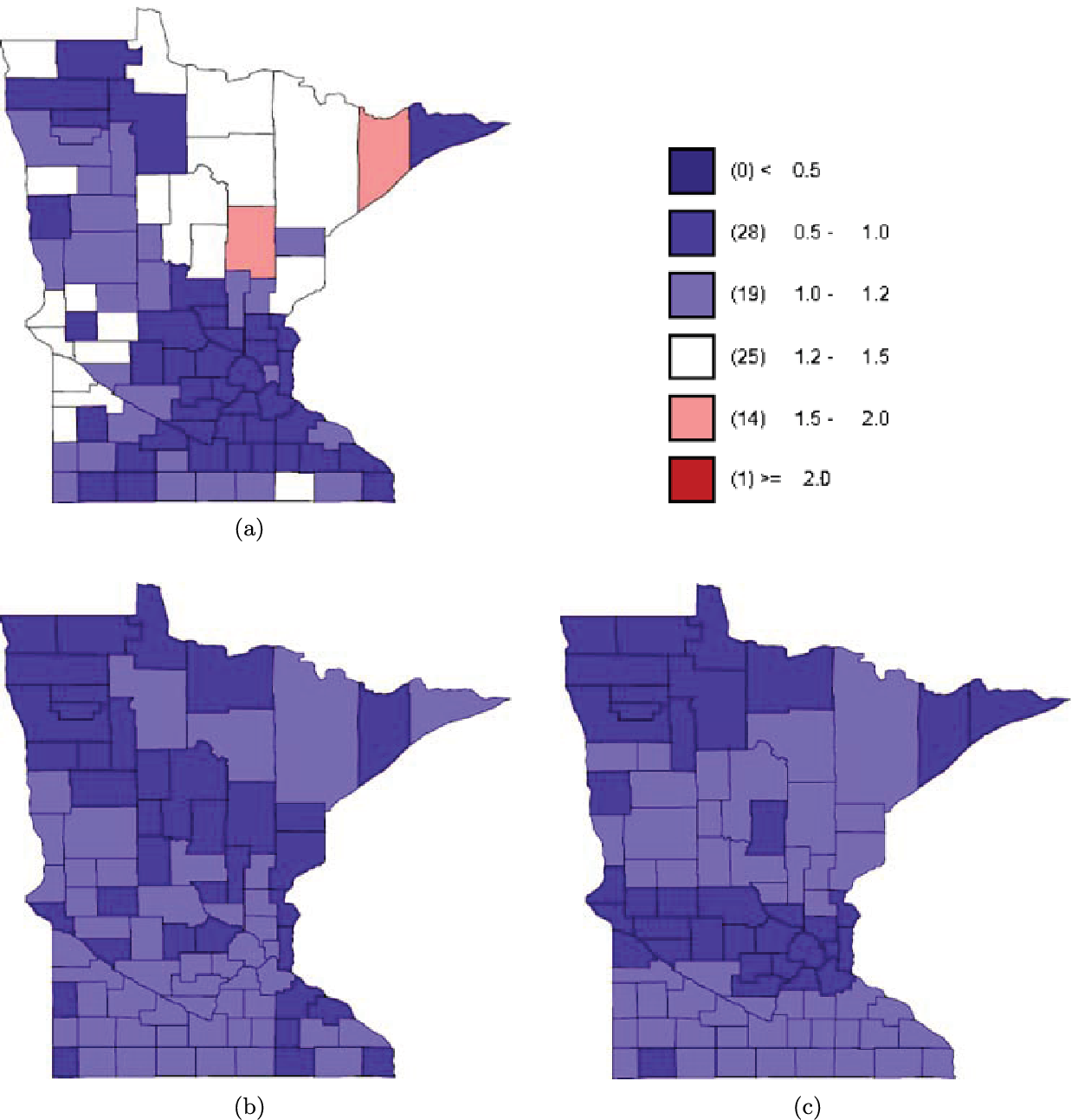}

\caption{SANOVA with $\mathit{HA}_1$: (\textup{a}) county main effect; (\textup{b}) cancer
$\times$
county interaction 1 for larynx; (\textup{c}) cancer $\times$ county interaction
2 for
esophagus.}
\label{m:main_int}
\end{figure}

To further examine the smoothing under SANOVA, Figure \ref{m:main_int}
shows separate maps for the county main effect and interactions from
the SANOVA fit with $\mathit{HA}_1$. The upper left plot is the cancer main
effect, the mean of the three cancers; the lower left plot is the
comparison of lung versus average of larynx and esophagus; the lower
right plot is the comparison of larynx versus esophagus. All values are
on the same scale as $y_{ij}/E_{ij}$ in Figures \ref{map1} to \ref
{map3} and use the same legend. The two interaction contrasts are
smoothed much more than the county main effect, agreeing with previous
research that larynx and esophagus cancer are closely related spatially
[Baron et al. (\citeyear{Baron93})]. To see whether the interactions
are necessary,
we fit a SANOVA model (using $\mathit{HA}_1$) without the interactions. As
expected, model complexity decreased ($p_D = 77$), while $\overline D$
increased slightly, so DIC became 1558, a bit better than SANOVA with
interactions.
Now consider the posterior of the MCAR's precision matrix $\Omega$. The
posterior mean of $\Omega$ is much larger for MCAR0.002 than MCAR200; the
diagonal elements are larger by 4 to 5 orders of magnitude. This may
explain the
poor coverage for MCAR0.002 in the simulation. Further, consider the correlation
matrix arising from the inverse of $\Omega$'s posterior mean. As the diagonals
of $R$ change from 0.002 to 200, the correlation between any two cancers
decreases and the complexity penalty $p_D$ increases. By comparison,
the three
SANOVAs have similar model fits and complexity penalties, leading to similar
DICs. So again, in this sense SANOVA shows greater
stability.\looseness=1

\begin{table}
\caption{Model comparison using $\mathit{DIC}$} \label{DIC}
\begin{tabular*}{270pt}{@{\extracolsep{\fill}}lccc@{}}
\hline
\textbf{Model} & $\bolds{\overline{D}}$ & $\bolds{p_D}$ & $\bolds{\mathit{DIC}}$\\
\hline
SANOVA with $\mathit{HA}_1$&1461&103&1564\\
SANOVA with $\mathit{HA}_2$&1463&102&1565\\
SANOVA with $H_A$ estimated from MCAR1&1458&103&1561\\
MCAR0.002 &1476&\phantom{0}79&1555\\
MCAR1 &1459&132&1591\\
MCAR200 &1559&356&1915\\
CAR0.001 &1461&141&1602\\
CAR1 &1453&149&1602\\
CAR1000 &1432&180&1612\\
\hline
\end{tabular*}
\end{table}

\section{\texorpdfstring{Discussion and future work.}{Discussion and future work}} \label{discussion}
We used SANOVA to do spatial smoothing and compared it with the much
more complex
MCAR model. For the cases considered here, we found SANOVA with spatial
smoothing to be an excellent
competitor to MCAR. It yielded essentially indistinguishable inference,
while being easier to fit and interpret. In the SANOVA model,
${H^{(+)}_A}$ is assumed known. For
most of the SANOVA models considered, ${H^{(+)}_A}$'s first column was
fixed to
represent the average over diseases, while other columns were
orthogonal to the
first column. Alternatively, ${H^{(+)}_A}$ could be treated as unknown and
estimated as part of the analysis. With this extension, SANOVA with spatial
effects is a reparameterization of the MCAR model and gains the MCAR model's
flexibility at the price of increased complexity. This extension would
be nontrivial, involving sampling from the space of orthogonal matrices
while avoiding
identification problems arising from, for example, permuting columns of
${H^{(+)}_A}$.
Other covariates can be added to a spatial SANOVA. Although (\ref
{long}) is a
saturated model, spatial smoothing ``leaves room'' for other
covariates. Such
models would suffer from collinearity of the CAR random effects and the fixed
effects, as discussed by Reich, Hodges and Zadnik (\citeyear{Reich06}), who gave a variant
analysis that
avoids the collinearity.
For data sets with spatial and temporal aspects, for example, the 11
years in the
Minnesota 3-cancer data, interest may lie in the counts' spatial
pattern and in
their changes over time. By adding a time effect, SANOVA can be
extended to a
spatiotemporal model. Besides spatial and temporal main effects, their
interactions can also be included and smoothed. There are many modeling choices;
the simplest model is an additive model without space-time
interactions, where
the spatial effect has a CAR model and the time effect a random walk,
which is a
simple CAR. But many other choices are possible.
We have examined intrinsic CAR models, where $Q_{ij}=-1$ if region $i$ and
region $j$ are connected. SANOVA with spatial smoothing could be
extended to
more general CAR models. Banerjee, Carlin and Gelfand (\citeyear{Banerjee04})
replaced $Q$ with the matrix
$D_w-\rho W$, where $D_w$ is diagonal with the same diagonal as $Q$ and
$W_{ij}=1$ if region $i$ is connected with region $j$, otherwise $W_{ij}=0$.
Setting $\rho=1$ gives the intrinsic CAR model considered in this
paper. For
known $\rho$, the SANOVA model described here is easily extended by replacing
$Q$ in Section 2 with $D_w-\rho W$. However, for unknown $\rho$, our method
cannot be adjusted so easily, because updating $\rho$ in the MCMC
would force
$V$ and the design matrix to be updated as well, but this would change the
definition of the parameter~$\bolds{\Theta}$. Therefore, a different
approach is
needed for unknown $\rho$.
A different extension of SANOVA would be to survival models for areal spatial
data [e.g., Li and Ryan~(\citeyear{Li02}), Banerjee, Wall and Carlin
(\citeyear{Banerjee03}), Diva,
Dey and Banerjee (2008)]. If the regions
are considered strata, then random effects corresponding to nearby
regions might
be similar. In other words, we can embed the SANOVA structure in a
spatial
frailty model. For example, the Cox model with SANOVA structure for
subject $j$
in stratum $i$ is
%
\begin{equation}
h(t_{ij},X_{ij}) = h_0(t_{ij}) \operatorname{exp}( X_{ij}\beta),
\end{equation}
where $X$ is the design matrix, which may include a spatial effect, a temporal
effect, their interactions and other covariates. Banerjee, Carlin and Gelfand
(\citeyear{Banerjee04}) noted
that in the CAR model, considering both spatial and nonspatial
frailties, the
frailties are identified only because of the prior, so the choice of
priors for
precisions is very important.
Besides the above extensions, HCSC introduced tools for normal SANOVA models
that can be extended to nonnormal SANOVA models. For example, HCSC
defined the
degrees of freedom in a fitted model as a function of the smoothing precisions.
This can be used as a measure of the fit's complexity, or a prior can
be placed
on the degrees of freedom as a way of inducing a prior on the unknowns
in the
variance structure. The latter is under development and will be
presented soon.
\section*{\texorpdfstring{Acknowledgments.}{Acknowledgments}}
The authors thank the referees, the Associate Editor and the Editor for
valuable comments and suggestions.

\begin{supplement} 
\stitle{Appendices, data and code}
\slink[doi]{10.1214/09-AOAS267SUPP}
\slink[url]{http://lib.stat.cmu.edu/aoas/267/supplement.zip}
\sdatatype{.zip}
\sdescription{Our supplementary material includes four sections as
appendices. In Appendix A we present a derivation of the precision
matrix of (\ref{precision}). Details of our MCMC algorithms can be
found in Appendix B. Appendix C discusses the mean transformation for
the Poisson case, while Appendix D discusses the estimation of the
$H^{(+)}_{A}$ from MCAR1. In addition, we provide a compressed folder
containing the data set for our 3-cancer Minnesota example as well as
an R code example to implement the SANOVA models.}
\end{supplement}

\printaddresses

\end{document}